\documentclass[fleqn]{article}

\usepackage{graphics}

\begin{document}

\title{ Coulomb effects on the transmittance of open \\ 
quantum dots in a tight-binding model} 

\author{ A.\ Aldea, A.\ Manolescu, and V.\ Moldoveanu}

\date{\empty}

\maketitle

\begin{center}
{\sl National Institute of Materials Physics,\\ 
P. O. Box MG-7 Bucharest-Magurele, Romania }
\end{center}

\begin{abstract}
A quantum-mechanical calculation of conductance in an open quantum
dot is performed in the Landauer-B\"{u}ttiker formalism using a
tight binding  Hamiltonian with direct Coulomb interaction.  The charge
distribution in the dot is calculated self-consistently as function of
a gate potential, for various dot-leads couplings.  The interaction
is active only inside the dot, but not in the leads, its strength
being an input parameter.  Our calculations are complementary to the
master-equation approach \cite{B}, go beyond the "orthodox theory", and
account for the size, tunneling, and interaction effects in quantum dots.
\end{abstract}

\section{Introduction}

It is known that the electron-electron interaction (EEI) is important in
the transport properties of small systems such as quantum dots (QD's).
The standard approach is the so-called "orthodox theory", which neglects
the size quantization and reduces the quantum mechanics to the charge
quantization inside a capacitor \cite{3}.  In this paper
we propose a pure quantum-mechanical method, based on the
Landauer-B\"{u}ttiker (LB) formalism and the Hartree approximation (HA).
We consider a discrete, tight-binding model, which allows the tailoring
of different shapes of the QD and the presence of a magnetic
field in the Peierls' fashion. The properties are controlled by several
inputs: a) the coupling of the QD with external leads,
$t_{LD}$;  b) the size and shape of the dot which determines
the electronic spectrum in the absence of the EEI;  c) the interaction 
strength, $U$. The calculations are  performed self-consistently in
the HA for dots of rectangular shape with four attached leads, 
accommodating up to 40 electrons.  The energy spectrum, transmittance,
and  dot charging are calculated and discussed. 

\section{The formalism}

We model the quantum dot as a 2D rectangular plaquette weakly
coupled to external leads.  In the tight-binding picture the Hamiltonian
consists of three terms, $H^{D}$ describing the isolated QD, $H^{L}$
for the leads, labeled by $\alpha$, and $H^{LD}$ for the coupling:
\begin{equation}
H=H^{D}+\sum_{\alpha}H_{\alpha}^{L} + \sum_{\alpha}H_{\alpha}^{LD} \,.
\end{equation}
Since  the role of the leads is only to inject and drain the electrons,
we neglect the EEI in $H^L$, and include it only in $H^D$. 
Meir and Wigreen and Gartner \cite{MW} have shown,
using different approaches, that in this situation the LB formalism remains
valid.

In this approach the Fermi level $E_F$ is fixed by the infinite leads, 
and the  external gate $V_g$ is simulated by the site energy in $H^{D}$:
\begin{equation}
H^{D}=\sum_{i} 
\Big(V_{g}+U\sum_{j \neq i}{<n_{j}>\over |i-j|}\Big)
\vert i\rangle\langle i\vert
+ t_{D}\sum_{<i,i'>}~\vert i\rangle\langle i'\vert \,.
\label{hdot}
\end{equation}
$<n_{j}>$ is the mean occupation number of the site $j\in$~QD, and $U$ is
the parameter describing the EEI.  We choose $t_{D}=1$ (i.e. the energy
unit is the hopping integral in QD) and denote by $<i,i'>$ the
nearest-neighbours summation.  
Instead of using the whole Hamiltonian (1), it is convenient to describe 
the open QD by an effective Hamiltonian with a non-hermitic term which 
depends on energy $E$ and incorporates the influence of the leads:
\begin{equation}
H_{eff}^{D}= H^{D}+
H^{DL}(E-H^{L})^{-1}H^{LD}= H^{D}+
\tau^2t_L\sum_{\alpha}~e^{-ik}~\vert\alpha\rangle\langle\alpha\vert \,,
\label{heff}
\end{equation}
where $t_L$ is the hopping energy of leads, 
$\tau=t_{LD}/t_L$,  $k$ is defined by $2t_L\cos k=E$, and $\vert\alpha\rangle$
stands for the dot state where the  lead $\alpha$ is attached. 

The matrix elements of the Green's function
$G^{+}_{ij}(E)=\langle i \mid (E-H_{eff}+i0)^{-1}\mid j \rangle $
are calculated numerically  using Eqs.(2-3) and the self-consistency condition
\begin{equation}
<n_{j}>={1\over\pi}\int_{-\infty}^{E_{F}} {\mathrm Im}~G_{jj}^{+}(E)\,dE\,.
\end{equation}
In the LB formalism the conductance 
matrix $g_{\alpha\beta}$ is given by the transmittance between 
the leads $\alpha$ and $\beta$:
\begin{equation}
g_{\alpha \beta}~=~{e^2\over h} T_{\alpha\beta}~
=4~{e^2\over h}\tau^4t_L^2{\sin^2}k~ |G_{\alpha\beta}^+(E_F)|^2 \,,
~~\alpha\not=\beta \,.
\label{1.2}
\end{equation}
In the weak-coupling limit $\tau\ll 1$, 
the transport problem is reduced
to a tunneling problem, and the transmittance shows resonances
corresponding to the energy spectrum of the isolated QD with the 
many-body effects included. The position, width and heights of the peaks 
depend on the above mentioned parameters.


\section{Results and discussion}

Each transmittance peak corresponds to the addition of an extra electron 
in a {\it Coulomb blockade} mechanism (for a review see Ref. 2). 
The position and width of resonances 
can be correlated to the Hartree spectrum of 
the isolated QD displayed in Fig.1 vs. $E_F$.  The  spectrum is
independent of $E_F$ as long as the later is not sufficiently close to an
energy level.
When $E_F$ approaches the $n$-th eigenvalue of
the system with $N$ electrons, and the addition of the $N+1$-th electron
becomes possible,
the whole spectrum
raises with the additional energy $E_{add}(N,N+1)= E_n(N+1)-E_n(N)$.
The essential characteristic is that the $E_{add}$  is not supplied
step-like but it is gained linearly with increasing $E_F$ in an
interval $\Delta E_F=E_{add}$ (in this region, a slope=1.0 is evident
in Fig.1).
This calculation shows that $E_{add}(N,N+1)$ depends on
the number of electrons $N$ and on the interaction
strength $U$. With increasing $U$, $E_{add}$ increases, the
plateaus of the energy levels shrink, while those of the 
number of particles increase.
A new electron is added exactly in the middle of the interval
$\Delta E_F$, i.e. at $E_F= (E_{n}(N)+E_{n}(N+1))/2$, 
see the diagonal line of Fig.1. These charge-degeneracy points 
correspond to the resonance peaks in Fig.2b.
The Coulomb blockade is accompanied by the so-called (elastic) {\it cotunneling}
effect meaning that the transmittance is non-vanishing 
also in a range about the degeneracy point \cite {Sch}. This effect 
gives rise to some width of the 
Coulomb oscillations. The comparison of Fig.1 and Fig.2b shows 
that the width {\em at the bottom of a conductance peak}
equals $E_{add}$, indicating that, for small $t_{LD}$, the cotunneling at
resonance is of Coulomb origin.
The number of electrons in the dot is a rather smooth function of the gate 
potential $V_{g}$, i.e. the steps of Fig.2a are not well defined 
even for a weak coupling (the dotted line),
in spite of the fact that the transmittance shows sharp peaks.
One notices that each of these peaks points
to an ideal step of Fig.2b, which corresponds to the limit $\tau \rightarrow 0$. 
For $\tau \sim 0.3-0.5$, the tunneling effects become more important: the
intra-valley cotunneling \cite{Averin} appears and even the resonance width
is now determined by tunneling. With still increasing the dot-leads coupling 
the resonances are spoiled, the charge quantization is totally lost and, 
if a strong enough magnetic field is applied, typical quantum 
Hall effects appear. We mention that here $t_{LD}$ is present in all powers 
of the perturbation theory.

\section{Conclusions}

This approach based on  Landauer-B\"{u}ttiker formalism, tight-binding
Hamiltonian and Hartree approximation is able to describe the interplay
of the size, interaction, and tunnelig in QD. Due to the size effects,
the addition energy  (in other words, the inverse of the dot capacitance)
depends on the number of electrons, and the Coulomb oscillations of
the conductance are distributed irregularly. Both intra-valley and
at-resonance cotunneling can be put into evidence; at resonance, a
competition between tunneling and interaction is present.  Increasing
the tunneling parameter, the charge quantization vanishes faster than the
resolution of the Coulomb peaks of the conductivity ; the peaks are 
enforced by the charge-degeneracy condition.

\vspace{1cm}

A.M. and A.A.
benefited from the hospitality and support of the International Centre for  
Theoretical Physics-Trieste at the Research Workshop on Mesoscopic Systems.
A.A. is grateful to Prof. Johannes Zittartz for his hospitality  at the
Institute of Theoretical Physics, University of Cologne, where the work
has been accomplished under SFB-341. Valuable discussions with Dr.P.Gartner are
acknowledged.

\newpage

\begin{figure}
\centerline{\resizebox{10cm}{6cm}{\includegraphics{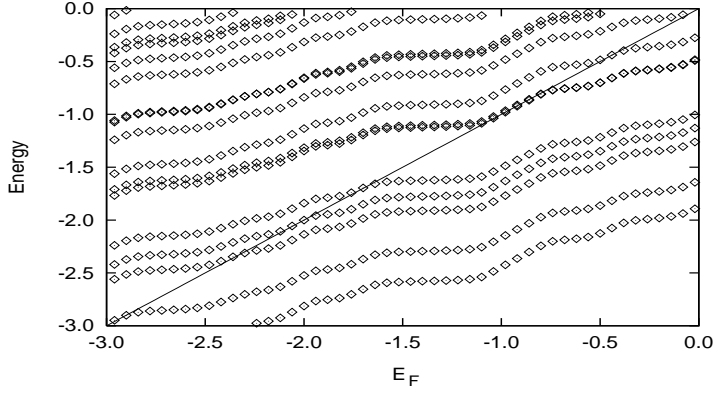}}}
\caption{The energy spectrum of the isolated QD vs. $E_F$, 
calculated self-consistently in the Hartree approximation for $U=0.5$. 
}
\label{fig1}
\end{figure}
\begin{figure}
\centerline{\resizebox{18cm}{10cm}{\includegraphics{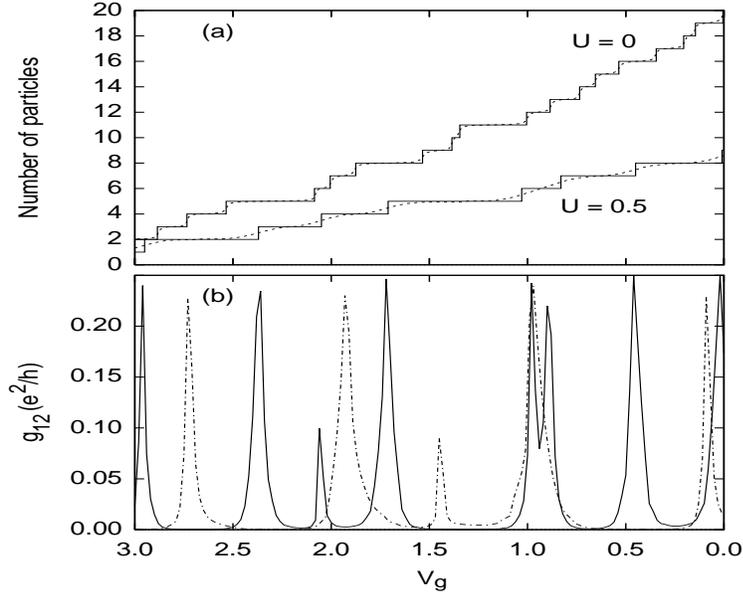}}}
\vspace{-2cm}
\caption{(a) The number of particles vs. the gate potential for $U=0$
(the noninteracting system) and for $U=0.5$.  With solid lines -
the limit $\tau\to 0$, with dashed lines - the weakly coupled system,
$\tau=0.1$.  (b) The conductance $g_{12}$ in the tunneling regime, for
$U=0.5$, with solid line, with maxima corresponding to the steps in the 
number of particles.  The dash-dotted line shows the conductance for $U=1$. }
\label{fig2}
\end{figure}

\end{document}